\documentclass[12pt]{article}
\usepackage{epsfig}
\usepackage{graphics}
\usepackage{graphicx}
\usepackage[centerlast,footnotesize]{caption2}
\begin{document}

\begin{center}
{\bf \Large  Production of Triply Charmed $\Omega_{ccc}$ Baryons \\
in $e^+e^-$ Annihilation}

\vspace{1.12cm}
S. P. Baranov$^{\it a}$\footnote{e-mail: baranov@sci.lebedev.ru} and 
V. L. Slad$^{\it b}$\footnote{e-mail: vslad@theory.sinp.msu.ru}

\vspace{0.75cm}

$^{\it a}${\small Lebedev Institute of Physics, Russian Academy of Sciences, 
Leninskii pr. 53, Moscow 119991, Russia}

$^{\it b}${\small Skobeltsyn Institute of Nuclear Physics, Moscow State 
University, Moscow 119992, Russia}
   
\vspace{1.35cm} 
                          
{\bf Abstract}
\end{center}\vspace*{-0.35cm}
{\small The total and differential cross sections for the production of triply 
charmed $\Omega_{ccc}$  baryons in $e^{+}e^{-}$ annihilation are calculated at 
the $Z$-boson pole.}

\vspace{1.02cm}

{\bf \center \Large 1. Introduction }

Investigation into the properties of baryons containing two or three heavy $c$ 
and $b$ quarks, the features
of their production at operating accelerators and
those under construction, and their lifetimes and
decay modes is topical in particle physics, but these
issues have not yet received adequate study. All that
is currently known in these realms from experiments
amounts to the claim [1] that a doubly charmed baryon $\Xi_{cc}^{+}$
was observed in experiments with a
beam of charged hyperons at FERMILAB. Theoretical investigations of baryons 
containing two heavy quarks are reviewed, for example, in [2]. Calculations
available in the literature that deal with the cross
sections for the production of baryons containing two
heavy quarks treat primarily processes described in
the fourth order of standard perturbation theory--that
is, processes leading to the production of respective
diquarks [3]. Only in [4] were sixth-order calculations
performed, where the process 
$e^+e^- \rightarrow s\bar{s} c\bar{c} b\bar{b} $
was
associated with the production of an $\Omega_{scb}$ baryon in
$e^+e^-$ collisions. The production of baryons involving
three heavy quarks has not yet been considered.

    The present article reports on a continuation of
the investigation begun in [4], providing a description of some features of the
process involving the production of triply charmed baryons $\Omega_{ccc}$ in 
$e^+e^-$ annihilation. This case has nothing to do with the
production $cc$ diquarks, because they can transform, with a 
probability close to unity, only into $\Xi_{ccu}^{++}$ or $\Xi_{ccd}^{+}$
baryons, a negligible fraction of these diquarks
going over to $\Omega_{ccc}$ baryons. As a matter of fact, calculations in the 
sixth order of perturbation theory for the elementary process 
$e^+e^- \rightarrow ccc\bar{c} \bar{c} \bar{c}$ are the only                  
possibility of theoretically studying triply charmed
baryons. The main contribution to the amplitude of
this process comes from 504 Feynman diagrams. In
relation to the production of $\Omega_{scb}$ baryons, which was
considered previously and where all components have
different flavors, the calculations for $\Omega_{ccc}$ baryons are
complicated by the need for taking into account the
interference between identical particles.
  
In studying the production of $\Omega_{ccc}$ baryons in
proton-proton collisions, it would be necessary to
consider an order of magnitude greater number of
Feynman diagrams corresponding to the subprocesses 
$q \bar{q} \rightarrow ccc\bar{c} \bar{c} \bar{c}  $  and  
$g g \rightarrow ccc\bar{c} \bar{c} \bar{c}  $. Moreover, the
description of baryon production in hadron-hadron
collisions would require a much greater effort in
calculating the contributions to the amplitude of the
production process from various parton color states
than in the case of $e^+e^-$ annihilation.
 
In present study, the unification of three charmed
quarks into an $\Omega_{ccc}$ baryon is described within the
well-known nonrelativistic approximation [5]. Upon
obtaining numerical results for the cross sections describing 
$\Omega_{ccc}$ baryon production, we analyze the possibility of constructing 
their approximate analytic description in terms of one known fragmentation 
function or another.

\vspace{1.02cm}

{\bf \center \Large 2. Amplitude of $\Omega_{ccc}$
production in  $e^+e^-$  annihilation}

We assume that the amplitude of the production
of triply charmed baryons $\Omega_{ccc}$ in $e^+e^-$ annihilation
corresponds to the elementary process
\begin{equation} 
e^+(k_{1}) + e^-(k_{2}) \rightarrow 
c(p_{1}, \xi_{1}) + c(p_{2}, \xi_{2}) + c(p_{3}, \xi_{2})
+ \bar{c}(p_{4}, \chi _{1}) + \bar{c}(p_{5}, \chi _{2}) + \bar{c}(p_{6}, 
\chi _{3}),  
\end{equation}
where $k_{1}$ and $k_{2}$ are the 4-momenta of colliding particles; 
$p_{1} ,...,  p_{6}$ are the 4-momenta of product partons; and $\xi_{i}$ and 
$\chi _{j}$ \  $(i,j=1,2,3)$  are the color indices
of quarks and antiquarks, respectively. As usual, we
disregard the contribution of the electroweak interaction of quarks to the 
amplitude of process (1), since
it is an order of magnitude less than the corresponding contribution of QCD 
interaction. Thereupon, all
Feynman diagrams to be taken into account for process (1) reduce to the nine 
basic diagrams in Fig.1,
which correspond to different positions of the quark-gluon vertices. Thirty-six
nonequivalent dispositions
of quark and antiquark lines characterized by specific 4-momenta, 
polarizations, and color indices are
possible for each of the basic diagrams {\it 1-7}, and {\it 18}
nonequivalent dispositions of such lines are possible
for the basic diagrams {\it 8} and {\it 9}. Since a collision between an 
electron and a positron leads to annihilation
into either a photon or a $Z$ boson, the total number of
relevant Feynman diagrams is 576.

First, we consider the color structure of the amplitude of $\Omega_{ccc}$ 
baryon production. Since an electron,
a positron, and any baryon are singlets with respect
to the $SU(3)_{c}$ color group, three product antiquarks c
must also form an  $SU(3)_{c}$-singlet state. Therefore,
the final state of process (1) must be fully antisymmetric in the color indices
of the three charmed
quarks bound into an $\Omega_{ccc}$ baryon and in the color
indices of the three product charmed antiquarks.
This requirement, together with the requirement of
an appropriate normalization, is satisfied by introducing, in the amplitude of 
process (1), the product
$(\varepsilon^{\xi_{1}\xi_{2}\xi_{3}}/\sqrt{6}) 
(\varepsilon^{\chi_{1}\chi_{2}\chi_{3}}/\sqrt{6})$
of antisymmetric tensors
and performing summation over the color indices of
$\xi_{i}$  and $\chi _{j}$  $(i,j=1,2,3)$.
  
We set $T^{a}=\lambda^{a}/2$, where $\lambda^{a}$  $(a=1,...,8)$ are
the Gell-Mann matrices, and denote by $N$ the total
number of such permutations of different pairs of
color indices of quarks and antiquarks that transform the sets  
$(\xi_{1}, \xi_{2}, \xi_{3} )$ and $(\chi_{1}, \chi_{2}, \chi_{3} )$ into the 
sets $(\xi_{i_{1},} \xi_{i_{2}}, \xi_{i_{3}} )$ and 
$(\chi_{i_{1}}, \chi_{i_{2}}, \chi_{i_{3}} )$, respectively. The color
factors associated with diagrams of the types {\it 1-7} can
then be found by means of direct analytic calculations.
The result is
\begin{equation}
\sum_{a,b,\zeta} \hspace{0.2cm} \sum_{\xi_{1}, \xi_{2}, \xi_{3}}
\hspace{0.2cm} \sum_{\chi_{1}, \chi_{2}, \chi_{3}}
\frac{1}{6} \varepsilon^{\xi_{1}\xi_{2}\xi_{3}}
\varepsilon^{\chi_{1}\chi_{2}\chi_{3}} T^{a}_{\xi_{i_{1}} \chi_{j_{1}}}
T^{a}_{\xi_{i_{2}} \zeta} T^{b}_{\zeta \chi_{j_{2}}}
T^{b}_{\xi_{i_{3}} \chi_{j_{3}}} =(-1)^{N} \frac{4}{9}.
\end{equation}

In the sum in expression (2), the index $\xi_{1}$, appears
twice (as it must)--directly in the tensor 
$\varepsilon^{\xi_{1}\xi_{2}\xi_{3}}$ and
indirectly as the substitute of one of the indices  
$\xi_{i_{1}}$, $\xi_{i_{2}}$ and $\xi_{i_{3}}$. The same is true for the other 
Greek indices in the above sum, with the exception of $\zeta$, and
for the zeroth color factors corresponding to diagrams
of types {\it 8} and {\it 9}, for which we have
\begin{equation}
\sum_{a,b,c} \hspace{0.2cm} \sum_{\xi_{1}, \xi_{2}, \xi_{3}}
\hspace{0.2cm} \sum_{\chi_{1}, \chi_{2}, \chi_{3}}
\varepsilon^{\xi_{1}\xi_{2}\xi_{3}}
\varepsilon^{\chi_{1}\chi_{2}\chi_{3}}
f^{abc} T^{a}_{\xi_{i_{1}} \chi_{j_{1}}}
T^{b}_{\xi_{i_{2}} \chi_{j_{2}}}
T^{c}_{\xi_{i_{3}} \chi_{j_{3}}} = 0,
\end{equation}
where $f^{abc}$ are the structure constants of the $SU(3)$
group. The proof of the equality in (3) is given in [4].
This equality means that the total contribution to the
amplitude of the process in (1) from the diagrams
involving three-gluon vertices vanishes. Thus, the
number of contributing diagrams reduces to 504.

Since the contribution to the amplitude of process
(1) from the diagram that differs from a specific diagram by a permutation of  
$N$ fermion pairs involves
the Feynman factor  $(-1)^{N}$, it can be concluded, with
allowance for (2), that all terms appearing in the
amplitude of $\Omega_{ccc}$ baryon production have the same sign.

It should be noted that, in our calculations, we
used an additional simplifying approximation, setting
the $c$-quark mass to zero in all expressions entering
into the numerators of fermion propagators and in all
traces. At the same time, we set $m_{c}=1.5$ GeV and
$p_{i}^{2}=2.25$ $\mbox{\rm GeV}^{2}$ $(i=1,...,6)$
in all of the denominators of the propagators of virtual particles and in the
expression for the final-state phase space (of course,
the amplitude containing zero $c$-quark mass in the
denominator would diverge). But if we used a nonzero
$c$-quark mass everywhere in the amplitude and in the
square of the relevant matrix element, the volume of
information to be saved in the computer memory and
the time required for the compilation of codes and
for numerical calculations of the cross sections would
grow enormously, which would render the problem in
question unsolvable with our means.

In order to estimate the effect of the above approximation on the accuracy of 
the numerical results,
we repeated the calculation of the cross section for
$\Omega_{scb}$ baryon production in a similar approximation
and compared the results obtained in this way with
the results of the full calculation performed in [4]. It
turned out that the cross sections obtained for $\Omega_{scb}$ baryon 
production within the "massive" and "massless" (for all quarks simultaneously) 
approximations differ only by 8$\%$. It seems reasonable to expect
an inaccuracy on the same order of magnitude for
$\Omega_{ccc}$ baryon production as well. Anyway, this inaccuracy does not 
exceed other theoretical uncertainties associated, for example, with the choice
of the renormalization scale in the strong-interaction coupling constant or 
with the wave function for the triply
charmed heavy baryon. Thus, this approximation appears to be numerically 
justified.

Taking into account the aforesaid, we can represent the matrix element for 
process (1) in the form
\begin{equation}
{\cal M}=\frac{g_{s}^{4}g^{2}}{9 \cos^{2}{\theta_{W}}  
(s-M^{2}_{Z}+iM_{Z}\Gamma_{Z})} D^{Z} - \frac{4g_{s}^{4}e^{2} }{9s}
D^{\gamma},
\end{equation}
where
\begin{eqnarray}
D^{Z}&=& \sum_{ i,j,k \in \{ 1,2,3 \}
\atop i \neq j \neq k} \sum_{i',j',k' \in \{1,2,3  \} 
\atop i' \neq j' \neq k'} \left\{                                                  
[({p}_{j}+{p}_{i}+{p}_{i'})^{2}-m_{c}^{2}]^{-1}               
[(k_{1}+k_{2}-{p}_{k'})^{2}-m_{c}^{2}]^{-1} \times \right. \\[3mm]
&\times&
({p}_{i}+{p}_{i'})^{-2}
({p}_{i}+{p}_{j}+{p}_{i'}+{p}_{j'})^{-2}              
\bar{u} ({\bf p}_{j}) \gamma^{\nu}                
(\hat{p}_{j}+\hat{p}_{i}+\hat{p}_{i'})                    
\gamma_{\delta} v(-{\bf p}_{j'}) \times   
            \nonumber  \\  
&\times&
\bar{u} ({\bf p}_{k}) \gamma^{\delta}                 
(\hat{k}_{1}+\hat{k}_{2}-\hat{p}_{k'}) \gamma_{\varepsilon} 
(g^{c}_{V}- g^{c}_{A}\gamma_{5}) v(-{\bf p}_{k'}) +  
            \nonumber  \\
&+&     
[({p}_{j'}+{p}_{i}+{p}_{i'})^{2}-m_{c}^{2}]^{-1}               
[(k_{1}+k_{2}-{p}_{k})^{2}-m_{c}^{2}]^{-1}            
({p}_{i}+{p}_{i'})^{-2} \times                                    
           \nonumber  \\
&\times& ({p}_{i}+{p}_{j}+{p}_{i'}+{p}_{j'})^{-2}              
\bar{u} ({\bf p}_{j})  \gamma_{\delta}                   
(-\hat{p}_{j'}-\hat{p}_{i}-\hat{p}_{i'})                   
\gamma^{\nu} v(-{\bf p}_{j'}) \times  
            \nonumber  \\  
&\times&
\bar{u} ({\bf p}_{k}) \gamma_{\varepsilon}                 
(-\hat{k}_{1}-\hat{k}_{2}+\hat{p}_{k}) \gamma^{\delta} 
(g^{c}_{V}- g^{c}_{A}\gamma_{5}) v(-{\bf p}_{k'}) +                      
            \nonumber  \\
&+&    
[({p}_{j'}+{p}_{i}+{p}_{i'})^{2}-m_{c}^{2}]^{-1}              
[(k_{1}+k_{2}-{p}_{k'})^{2}-m_{c}^{2}]^{-1}           
({p}_{i}+{p}_{i'})^{-2} \times                                    
           \nonumber  \\
&\times& ({p}_{i}+{p}_{j}+{p}_{i'}+{p}_{j'})^{-2}             
\bar{u} ({\bf p}_{j}) \gamma_{\delta}  
(-\hat{p}_{j'}-\hat{p}_{i}-\hat{p}_{i'}) \gamma^{\nu}  
v(-{\bf p}_{j'}) \times    
            \nonumber  \\  
&\times&
\bar{u} ({\bf p}_{k}) \gamma^{\delta}   
(\hat{k}_{1}+\hat{k}_{2}-\hat{p}_{k'}) \gamma_{\varepsilon} 
(g^{c}_{V}- g^{c}_{A}\gamma_{5}) v(-{\bf p}_{k'}) +                      
            \nonumber  \\
&+&     
[({p}_{j}+{p}_{i}+{p}_{i'})^{2}-m_{c}^{2}]^{-1}              
[(k_{1}+k_{2}-{p}_{k})^{2}-m_{c}^{2}]^{-1}           
({p}_{i}+{p}_{i'})^{-2} \times                                    
           \nonumber  \\
&\times& ({p}_{i}+{p}_{j}+{p}_{i'}+{p}_{j'})^{-2}             
\bar{u} ({\bf p}_{j}) \gamma^{\nu}               
(\hat{p}_{j}+\hat{p}_{i}+\hat{p}_{i'}) \gamma_{\delta}  
v(-{\bf p}_{j'}) \times  
            \nonumber  \\  
&\times&
\bar{u} ({\bf p}_{k}) \gamma_{\varepsilon} 
(-\hat{k}_{1}-\hat{k}_{2}+\hat{p}_{k}) \gamma^{\delta}       
(g^{c}_{V}- g^{c}_{A}\gamma_{5}) v(-{\bf p}_{k'})+   
            \nonumber  \\
&+&  
[({p}_{j}+{p}_{j'}+{p}_{k})^{2}-m_{c}^{2}]^{-1}              
[(k_{1}+k_{2}-{p}_{k'})^{2}-m_{c}^{2}]^{-1}           
({p}_{i}+{p}_{i'})^{-2} \times                                    
           \nonumber  \\
&\times& ({p}_{j}+{p}_{j'})^{-2}                         
\bar{u} ({\bf p}_{j}) \gamma_{\delta} v (-{\bf p}_{j'}) 
\bar{u} ({\bf p}_{k}) \gamma^{\delta}
(\hat{p}_{j}+\hat{p}_{j'}+\hat{p}_{k}) \times         
                \nonumber  \\  
&\times&
\gamma^{\nu} (\hat{k}_{1}+\hat{k}_{2}-\hat{p}_{k'})           
\gamma_{\varepsilon} (g^{c}_{V}- g^{c}_{A}\gamma_{5})
v(-{\bf p}_{k'}) +                                           
            \nonumber  \\
&+&     
[({p}_{j}+{p}_{j'}+{p}_{k})^{2}-m_{c}^{2}]^{-1}              
[({p}_{i}+{p}_{i'}+{p}_{k'})^{2}-m_{c}^{2}]^{-1}                   
({p}_{i}+{p}_{i'})^{-2}\times                                      
           \nonumber  \\
&\times& ({p}_{j}+{p}_{j'})^{-2}
\bar{u} ({\bf p}_{j}) \gamma_{\delta} v(-{\bf p}_{j'}) 
\bar{u} ({\bf p}_{k}) \gamma^{\delta}  
(\hat{p}_{j}+\hat{p}_{j'}+\hat{p}_{k}) \times 
                \nonumber  \\  
&\times&
\gamma_{\varepsilon} (-\hat{p}_{i}-\hat{p}_{i'}-\hat{p}_{k'}) 
\gamma^{\nu} (g^{c}_{V}- g^{c}_{A}\gamma_{5}) 
v(-{\bf p}_{k'})+                                            
            \nonumber  \\
&+&      
[({p}_{i}+{p}_{i'}+{p}_{k'})^{2}-m_{c}^{2}]^{-1}              
[(k_{1}+k_{2}-{p}_{k})^{2}-m_{c}^{2}]^{-1}           
({p}_{i}+{p}_{i'})^{-2}\times                                      
           \nonumber  \\
&\times& ({p}_{j}+{p}_{j'})^{-2}                                
\bar{u} ({\bf p}_{j}) \gamma_{\delta} v (-{\bf p}_{j'})
\bar{u} ({\bf p}_{k}) \gamma_{\varepsilon} 
(-\hat{k}_{1}-\hat{k}_{2}+\hat{p}_{k}) \times 
                \nonumber  \\  
&\times& \left.
\gamma^{\delta} (-\hat{p}_{i}-\hat{p}_{i'}-\hat{p}_{k'})
\gamma^{\nu} (g^{c}_{V}- g^{c}_{A}\gamma_{5})                            
v(-{\bf p}_{k'}) \right\} \times 
              \nonumber  \\
&\times&
\bar{u}({\bf p}_{i}) \gamma_{\nu} v(-{\bf p}_{i'}) 
\bar{v}(-{\bf k}_{1}) \gamma^{\varepsilon} 
(g^{e}_{V}-g^{e}_{A}\gamma_{5}) u({\bf k}_{2}),       
\nonumber           
\end{eqnarray}
while the expression for $D^{\gamma}$ can be derived from $D^{Z}$
by means of the substitutions 
$g^{e}_{V} \rightarrow 1$, $g^{e}_{A} \rightarrow 0$,     
$g^{c}_{V} \rightarrow Q_{c}=2/3$  and $g^{c}_{A} \rightarrow 0$.
Summation in (5) corresponds to 36 permutations of quark and antiquark
lines in diagrams of types {\it 1-7}.

\vspace{1.02cm}

{\bf \center \Large 3. Method of orthogonal amplitudes}

In order to derive the expression that is obtained
for the square of the matrix element $(\overline{|{\cal M}|^{2}})$ upon
summation over the final-fermion polarizations and
averaging over the polarizations of colliding particles,
we use the method of orthogonal amplitudes and the
REDUCE computer system for analytic calculations.
The method of orthogonal amplitudes was proposed
in [6] and was employed in calculations referring to
$\Omega_{scb}$ baryon production in $e^+e^-$ collisions [4].

A simple and mathematically rigorous validation
of the method of orthogonal amplitudes is the following (to the best of our 
knowledge, it has not yet
been given anywhere). Suppose that four-component
spinors $u(\bf p)$ and $u(\bf p')$ describing particles of mass
$m$ and  $m'$, respectively, their 4-momenta being $p$
and $p'$ ($p^{2}=m^{2}$, $p'^{2}=m'^{2}$), obey the Dirac equation.
Of the four linear homogeneous equations for the
components of the spinor $u(\bf p)$ [$u(\bf p')$], only two are
independent; therefore, each of the four components
under consideration can be represented as a linear
combination of two arbitrary independent constants,
denoted here by $X$ and $Y$ ($X'$ and $Y'$). Any quantity of
the form $\bar{u}({\bf p'}) R u({\bf p})$, where $R$ is an operator specified
in terms of the $\gamma$-matrices and their contractions with
some 4-vectors, can be represented as a linear combination of four independent 
elements $XX'^{*}$, $XY'^{*}$, $YX'^{*}$, and $YY'^{*}$. 
Therefore, quantities of the form
$\bar{u}({\bf p'}) R u({\bf p})$
can be treated as vectors of a linear four-dimensional space $L$ spanned by 
the above elements. Any four linearly independent quantities of the form
$w_{n} \equiv \bar{u}({\bf p'}) O_{n} u({\bf p})$, where the operator $O_{n}$ 
is either unity, $\gamma_{5}$, $\hat{V}$,  
$\hat{V'}\gamma_{5}$, or $(\hat{V''}\hat{V'''}-\hat{V'''}\hat{V''})/2$ (with 
$V$, $V'$, $V''$,  and $ V'''$ being arbitrary 4-vectors), can be
taken for basis vectors of the space $L$. The scalar product 
$(w_{n}, w_{n'})$ of vectors $w_{n}$ and $w_{n'}$ belonging
to the linear space $L$ is defined as the product $w_{n} w_{n'}^{*}$
summed over the polarizations of fermions that are
described by the spinors $u({\bf p})$ and $u({\bf p'})$.

Here, we take, for basis vectors of the space $L$,
four quantities $w_{n}$ specified by the operators  $O_{1} = 1$, 
$O_{2} = \hat{K}$, $O_{3} = \hat{Q}$, and  $O_{4} = \hat{K}\hat{Q}$ with the 
4-vectors $K$ and   $Q$   here being orthogonal to the 4-momenta $p$ and $p'$ 
and to each other - that is, 
$K_{\mu}p^{\mu}=0$, $K_{\mu}p'^{\mu}=0$, $Q_{\mu}p^{\mu}=0$, 
$Q_{\mu}p'^{\mu}=0$, and $K_{\mu}Q^{\mu}=0$.
Otherwise, the 4-vectors $K$ and $Q$ are arbitrary.
They can be specified, for example, by the relations 
$K^{\mu}$ = $\varepsilon^{\mu\nu\rho\sigma}p_{\nu}p'_{\rho}a_{\sigma}$ and 
$Q^{\mu}$ = $\varepsilon^{\mu\nu\rho\sigma}p_{\nu}p'_{\rho}K_{\sigma}$, where
the 4-vector $a_{\sigma}$ is entirely arbitrary. From the orthogonality of the 
4-vectors  $K$ and   $Q$, it follows that 
$\hat{K}\hat{Q}$ = $(\hat{K}\hat{Q}-\hat{Q}\hat{K})/2$.
The four quantities $w_{n}$ used
are orthogonal to one another, $(w_{n}, w_{n'})$ = $C_{n}\delta_{nn'}$,
$C_{n} \neq 0$, this proving their linear independence and
justifying the name "orthogonal amplitudes." Thus, it
was shown that any quantity of the form $\bar{u}({\bf p'}) R u({\bf p})$
can be represented as a linear combination of orthogonal amplitudes.

In order to solve the problem of calculating the
square of the matrix element, we first introduce basic
orthogonal amplitudes as
\begin{equation}
w_{i1}=\bar{u}({\bf p}_{i}) v(-{\bf p}_{3+i}), \hspace{0.5cm} 
w_{i2}=\bar{u}({\bf p}_{i})\hat{K}_{i}  v(-{\bf p}_{3+i}),
\end{equation}
$$w_{i3}=\bar{u}({\bf p}_{i})\hat{Q}_{i} v(-{\bf p}_{3+i}), \hspace{0.5cm}  
w_{i4}=\bar{u}({\bf p}_{i})\hat{K}_{i}\hat{Q}_{i}  v(-{\bf p}_{3+i}),$$
$$w_{e_{1}}=\bar{v}(-{\bf k}_{1}) u({\bf k}_{2}), \hspace{0.5cm}
w_{e_{2}}=\bar{v} (-{\bf k}_{1}) \hat{K}_{e} u({\bf k}_{2}),$$
$$w_{e_{3}}=\bar{v}(-{\bf k}_{1}) \hat{Q}_{e} u({\bf k}_{2}), \hspace{0.5cm}
w_{e_{4}}=\bar{v} (-{\bf k}_{1}) \hat{K}_{e}\hat{Q}_{e} u({\bf k}_{2}),$$
where $i=1,2,3$. We would like to note that the pair
combinations of the spinors $\bar{u}({\bf p}_{i})$ and  $v(-{\bf p}_{j})$ can 
be chosen in six equivalent ways.

On the basis of the quantities in (6), we construct
256 orthogonal amplitudes as
\begin{equation} 
w_{nrst}=w_{1n}w_{2r}w_{3s}w_{et},
\end{equation}
where $n,r,s,t = 1,2,3,4$.

The expansion of the matrix element (4) in the
amplitudes given by (7) has the form
\begin{equation}
{\cal M}=\sum \limits_{n,r,s,t=1}^{4} c_{nrst} w_{nrst}. 
\end{equation}

In order to derive the coefficients $c_{nrst}$ in this expansion, we multiply 
both sides of (8) by the factor
$w_{n'r's't'}$, take the sum of the result over the polarizations of all of the
fermions, and make use of the
orthogonality of different amplitudes. As a result, we arrive at
\begin{equation} 
c_{nrst}= \{ \sum_{polar.} {\cal M} w_{nrst}^{*} \}/(w_{nrst},w_{nrst}),
\end{equation}
where $(w_{nrst}, w_{nrst})$ is an analog of the scalar product
defined above in the linear space $L$ - that is, the sum
of the squared modulus of the amplitude $w_{nrst}$ over
the polarization of all fermions. Since $w_{nrst}$ involves
arbitrariness associated with the choice of the 4-
vectors $K$ and $Q$ in (6), there is also arbitrariness
in the coefficients $c_{nrst}$ in (9). The substitution of
these coefficients into (8) leads to an identity whose
left-hand side is determined unambiguously. Thus,
summation on the right-hand side of (8) removes the
above ambiguity.

Since electrons and positrons are treated as massless particles and since the 
charmed-quark mass is
set to zero in the numerators of each term of the
amplitude for process (1) and in respective traces, it is
clear that 192 of the 256 coefficients in expansion (8)
vanish, because they are linear combinations of the
traces of an odd number of the Dirac  $\gamma$ matrices. We
further list 64 orthogonal amplitudes in formula (7)
that generate nonzero expansion coefficients: $t=2,3$
with either       $n,r,s=2,3$, or one of the indices $n,r,s$ is
equal to 2 or 3, while the other two belong to the set
$\{1, 4\}$.

It can clearly be seen that the expression obtained
for the square of the matrix element upon summation
over the polarizations of final fermions and averaging
over the polarizations of initial particles takes the form
\begin{equation}
\overline{|{\cal M}|^{2}}=
\frac{1}{4} \sum \limits_{n,r,s,t} |c_{nrst}|^2 \cdot (w_{nrst},w_{nrst}).
\end{equation}

In actual calculations by the method of orthogonal
amplitudes, we compose one REDUCE code for
traces and tensor contractions that corresponds to
504 terms in any quantity ${\cal M}w^{*}_{nrst}$ and then, by
means of any text editor (for example, "joe"), perform
obvious changes necessary for obtaining the REDUCE code for calculating all 64 
nonzero quantities ${\cal M}w^{*}_{nrst}$.

\vspace{1.02cm}

{\bf \center \Large  4. Cross sections  for $\Omega_{ccc}$ baryon production
at the $Z$ pole in $e^{+}e^{-}$ annihilation}

In describing the $\Omega_{ccc}$ baryon as a bound state
of three charmed quarks, we use the nonrelativistic
approximation [5]. This means that we disregard the
relative velocities of the $c$ quarks confined within the
baryon - that is, in the laboratory frame, the velocities
and momenta of all three $c$ quarks produced in process
(1) are taken to be identical and equal to one-third of
the momentum $p$ of the $\Omega_{ccc}$ baryon having the mass
$M$=$3m_{c}$. With allowance for the unification of three
charmed quarks into the baryon, the phase space of
process (1) effectively becomes the 4-particle phase
space of the process
\begin{equation} 
e^+(k_{1}) + e^-(k_{2}) \rightarrow \Omega_{ccc}(p)
+ \bar{c}(p_{4}) + \bar{c}(p_{5}) + \bar{c}(p_{6}).
\end{equation}

The differential cross section for process (11) takes the form
\begin{equation}
d\sigma = \frac{  (2\pi)^{4} \overline{|{\cal  M}|^{2} } }{ 2 s } \cdot  
\frac{ |\psi(0)|^{2} }{ M^{2} }
\delta^{4}( k_{1}+k_{2}- p - p_{4}- p_{5}- p_{6} ) \times 
\end{equation} 
$$\times \frac{d^{3}p}{ (2\pi)^{3} 2E} \cdot
\frac{d^{3}p_{4}}{ (2\pi)^{3} 2E_{4}} \cdot 
\frac{d^{3}p_{5}}{ (2\pi)^{3} 2E_{5}} \cdot
\frac{d^{3}p_{6}}{ (2\pi)^{3} 2E_{6}},$$
where  $\psi(0)$ is the value that the respective wave function takes in the 
case where all three $c$ quarks forming
the $\Omega_{ccc}$ baryon are located at the same point, so
that their relative coordinates are zero. The numerical
value of $|\psi(0)|^{2}$  is taken to be identical to that in [7],
where it was
\begin{equation} 
|\psi(0)|^{2} = 0.36 \cdot 10^{-3} \hspace{0.2cm} \mbox{\rm GeV}^{6} .
\end{equation}  

In calculating the total and differential cross sections, we employed codes for
numerical integration
that are based on the Monte Carlo method and which
are contained in the CompHEP package [8], which
is broader. It appeared that the maximum computational errors in the 
differential cross sections came
from the first iteration. Therefore, only the total cross
section was calculated in the first iteration, while
both the total cross section and the differential cross
sections were determined in the next five iterations.
Each iteration involved 200 000 Monte Carlo calls
on the integrand. The errors in calculating the total
cross section amounted to 1.0\%, while the errors in
calculating the differential cross sections were predominantly 2 to 3\% 
(this is reflected below in the text
and in the figures). As was indicated above, the error
associated with the disregard of the charmed-quark
mass in the numerators of the amplitude for process
(1) and in the traces is a few percent. Moreover,
we additionally tested the consistency of the cross-section values for two 
different admissible choices of the 4-vectors      
$K^{\mu}_{e}, Q^{\mu}_{e}, K^{\mu}_{i}$, and $Q^{\mu}_{i}$ $(i=1,2,3)$  used
to construct the quantities in (6), which specify the
orthogonal amplitudes (7).

In addition to statistical errors, the calculations
contain unavoidable theoretical uncertainties. First,
there is the uncertainty associated with the running
strong-interaction coupling constant as a function of
the renormalization scale. Since all of the calculations
were performed at an energy value that corresponds
to the $Z$-boson pole ($\sqrt{s}$ = 91.2 GeV), it is reasonable to specify the 
coupling-constant values as
follows: $\alpha_{s}$ = $\alpha_{s}(M_{Z}/2)$  = 0.134 and  
$\alpha$ = $\alpha (M_Z )$ = 1/128.0; 
accordingly, sin$^2\theta_W$ = sin$^2\theta_W (M_Z )$ = 0.2240. 
However, it is not evident why it is $M_Z$,
and not, for example, the invariant mass of some
product quark pair or even the $\Omega_{ccc}$  baryon mass,
that should be chosen for the characteristic scale of
strong interaction. Since the cross section for process
(11) is proportional to the fourth power of the strong-
interaction coupling constant, this source of errors
is the most important. Second, the accuracy of the
potential model employed as a basis for calculating
the baryon-wave-function value $\psi (0)$ is uncertain.

For the chosen set of model parameters, the total
cross section for the process $\sigma_{\rm tot}$ and the forward
(backward) production asymmetry at the $Z$-boson pole are
\begin{equation}
\sigma_{\mbox{\rm \small tot}}=(0.0404 \pm 0.0004) \hspace{0.2cm} 
\mbox{\rm fb},
\end{equation}
\begin{equation}
A_{FB}=(\sigma_{F} - \sigma_{B})/(\sigma_{F} + \sigma_{B})=
0.101 \pm 0.005,
\end{equation}
where $\sigma_{F} (\sigma_{B}) $ is the cross section for the production
of an $\Omega_{ccc}$ baryon moving in the forward (backward)
direction with respect to the electron-momentum direction. The cross-section 
value in (14) is close to
that of the total cross section for $\Omega_{scb}$ baryon production in 
$e^+e^-$ collisions (0.0534 $\pm$ 0.0014 fb) if the
strange-quark mass is set to 300 MeV [4].

The differential cross sections with respect to the
transverse momentum $p_T$ and the rapidity $Y$ of $\Omega_{ccc}$
baryons are presented in Fig. 2. The distribution
$d\sigma/dY$ peaks at a small positive value of $Y$, while
$d\sigma / dp_{T}$ has a maximum at $p_T$ $\approx$ 12 GeV. Note that
the maximum of the differential cross section with
respect to the transverse momentum of $\Omega_{ccc}$ baryons
occurs at a $p_T$ value much lower than that for $\Omega_{scb}$
baryons produced under the same conditions, in
which case $d\sigma / dp_{T}$ peaks within the $p_T$ interval 23-26 GeV.

It is desirable to associate our numerical results
with some simple analytic form, which we will seek
among well-known fragmentation functions [9-12].
It is clear that the production of a triply charmed
baryon can hardly be interpreted as a fragmentation
process, since each of the three $c$ quarks can be
treated, on equal footing, as a fragmenting quark
produced at the  $\gamma /Z$ vertex and since the interference
between identical quarks is likely to be significant in
the process being considered. However, we accept not
the physical concept of the fragmentation model but
its mathematical form used in processing experimental data on 
$e^+e^-$ annihilation (see, for example, [13]);
namely, we set
\begin{equation}
\frac{d\sigma}{dz}
=\sigma_{c\bar{c}} \cdot D_{c \rightarrow \Omega_{ccc}} (z),
\end{equation}
where $\sigma_{cc}$ is the total cross section for the process
$e^{+}e^{-}  \rightarrow  c\bar{c}$
while $D_{c \rightarrow \Omega_{ccc}} (z)$ is the respective fragmentation 
function. Instead of the variable $z$, its approximate value 
$x_p = p/p_{\rm max}$ is used below.

Neglecting a small asymmetry in the angular distribution of $\Omega_{ccc}$ 
baryons, we arrive at the following
relation between the differential cross section with
respect to the transverse momentum and the fragmentation function:
\begin{equation}
\frac{d\sigma}{dp_{T}}
=\frac{4 \sigma_{c\bar{c}} p_{T}}{s}
\int \limits^{1}_{2 p_{T} / \sqrt{s}} 
\frac{D_{c \rightarrow \Omega_{ccc}}(z) dz}
{z \sqrt{z^{2} - 4 p_{T}^{2}/s}}.
\end{equation}

We now compare our numerical results with those
obtained according to expression (17) with various
fragmentation functions. First, we consider the Peterson function [9]
\begin{equation}
D(z) \sim \frac{1}{z} \left( 1- \frac{1}{z}
- \frac{\varepsilon}{1-z}\right)^{-2}, 
\end{equation}
which is often used in processing experimental data
on charmed-hadron production in $e^+e^-$ annihilation
[14]. Also, this function provides a good approximation to numerical results 
on $\Omega_{scb}$ baryon production
in $e^+e^-$ annihilation [4]. The best fit to our calculations (dash-dotted 
curve in Fig. 2) corresponds to
$\varepsilon= 0.92$ . The agreement is clearly poor. The best fits
with the Collins-Spiller fragmentation function [10]
\begin{equation}
D(z) \sim \left( \frac{ 1 - z }{ z } + \varepsilon \frac{ 2 - z }{ z } \right)
( 1 + z^{2}  )\left( 1- \frac{1}{z} - \frac{\varepsilon}{1-z}\right)^{-2} 
\end{equation}
at  $\varepsilon=3.0$ and with the fragmentation function [11]
\begin{equation}
D(z) \sim z^{\alpha} ( 1 - z ) , 
\end{equation}        
at $\alpha=0.8$ are also unsatisfactory. The results of the
calculations according to (17) with the functions in
(19) and (20) are displayed in Fig. 2 (dashed and
dotted curves, respectively).

An acceptable analytic form for our numerical results is provided by the LUND 
fragmentation function [12]
\begin{equation}
D(z) \sim \frac{1}{z}
( 1 - z )^{a}
\mbox{\rm exp}\left( -\frac{c}{z} \right) , 
\end{equation}
at the parameters $a=2.4 \pm 0.2$ and $c=0.70 \pm 0.03$.
The corresponding results calculated according to
(17) are represented by the solid curve in Fig. 2.

It should be noted that the fragmentation functions (19)-(21), along with the 
functions in (18), were employed by the OPAL Collaboration [15] in
processing experimental data on $B$-meson production.

\newpage

\begin{figure}[ht]
\vspace*{-5.7cm}  \hspace*{-2.9cm}
\centering \includegraphics[scale=0.91]{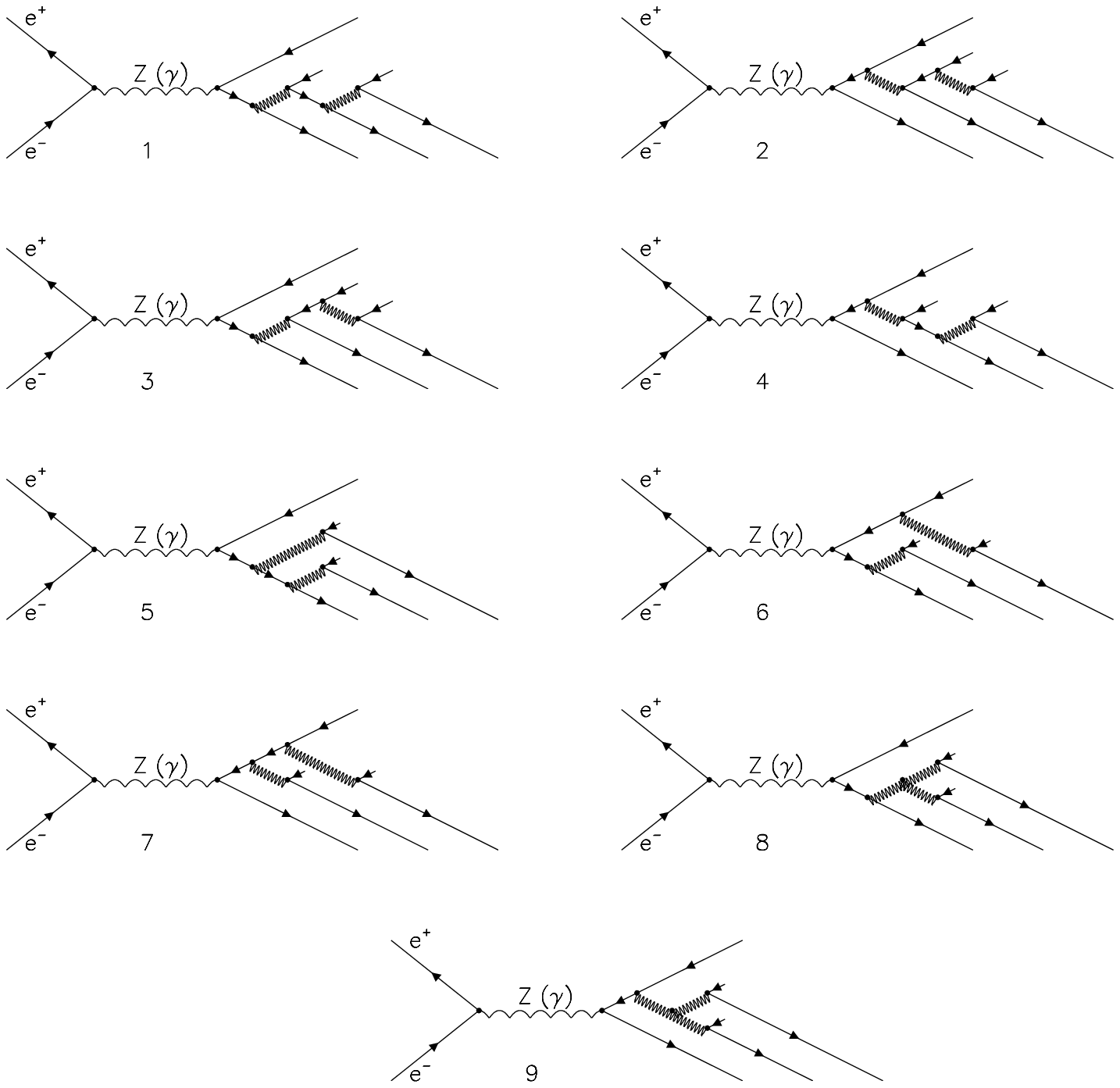}
\vspace*{-7.8cm}
\end{figure}
\begin{small}
{\footnotesize{\bf \hspace*{0.3cm} Fig. 1.} Basic Feynman diagrams for the 
process $e^{+}+e^{-} \rightarrow c+c+c +\bar{c} +\bar{c}+ \bar{c}$.}
\end{small}

\newpage

\begin{figure}[ht]
\centering \includegraphics[width=7.9cm]{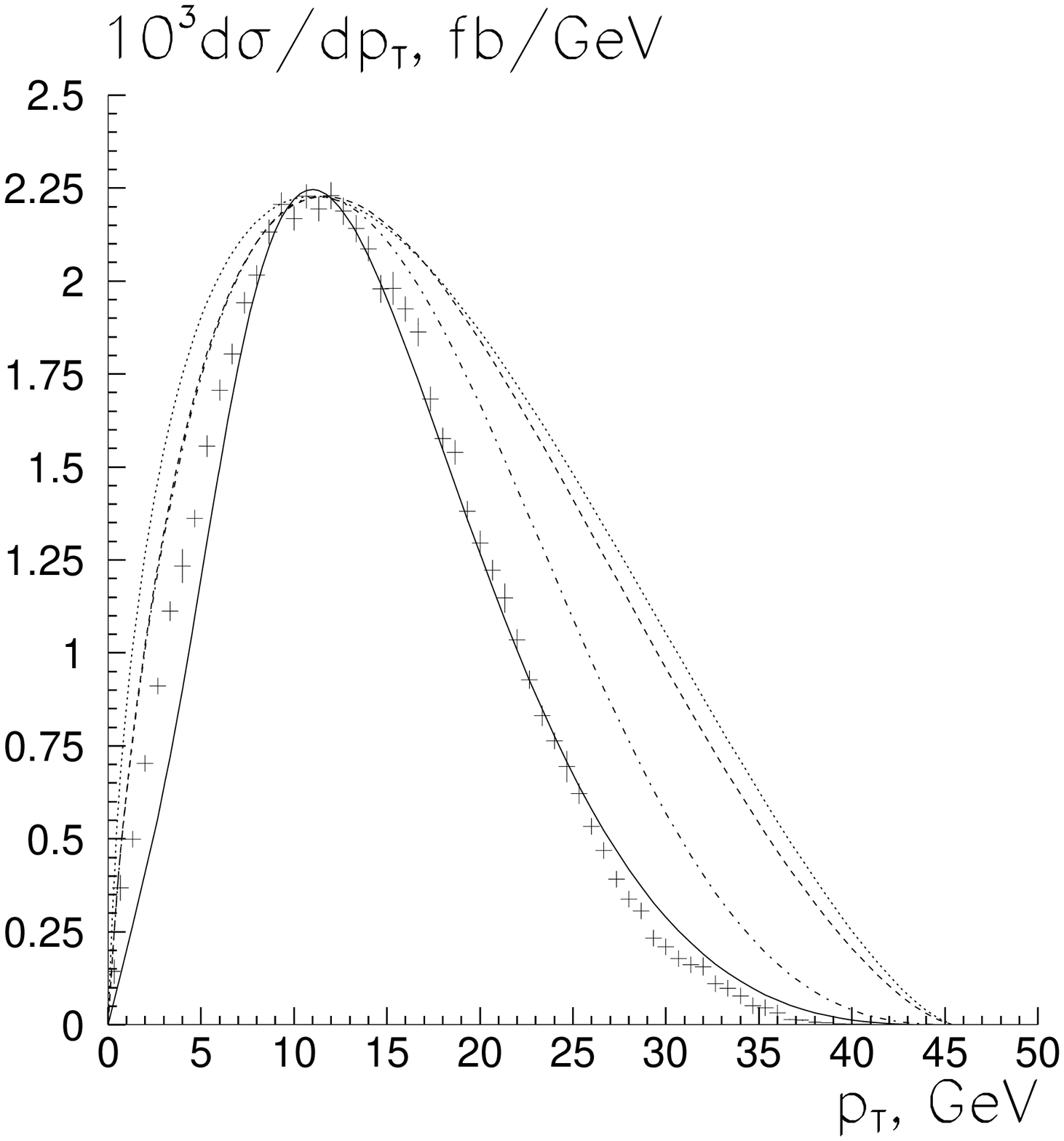} 
\includegraphics[width=7.9cm]{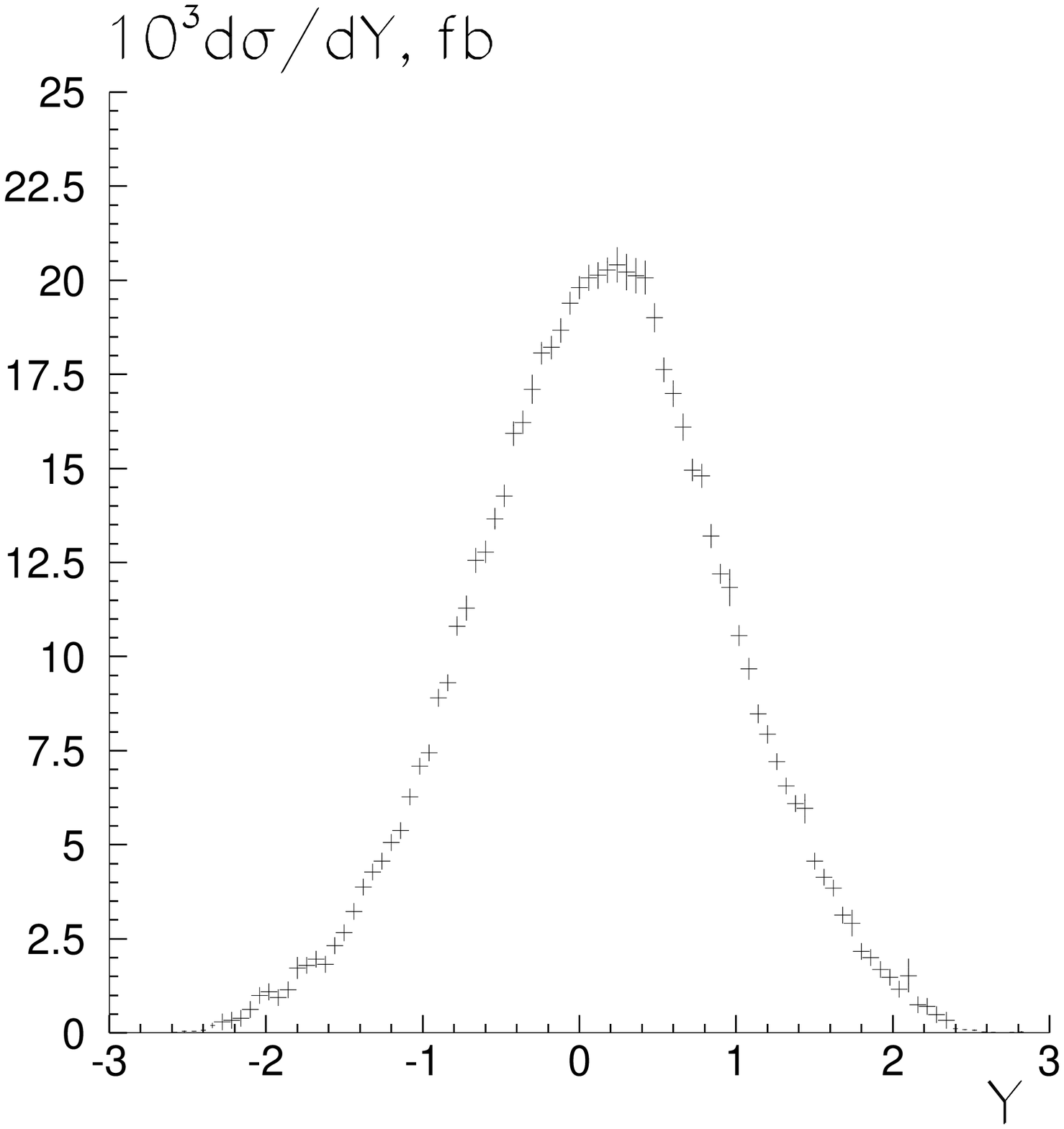}
\end{figure}

\begin{footnotesize}
{\bf Fig. 2.}  
Differential cross sections for $\Omega_{ccc}$ baryon production in $e^+e^-$ 
annihilation at the $Z$ pole with respect to the 
transverse momentum $p_T$ (left) and  rapidity $Y$ (right). The results of 
Monte Carlo calculations and the errors in them are represented
by crosses. The curves in Fig. 2 (left) correspond to expression (17) 
calculated with the fragmentation functions in the form (solid
curve) (21) at $a$ = 2.4 and $c$ = 0.70, (dash-dotted curve) (18) at  
$\varepsilon$ = 0.92, (dashed curve) (19) at  $\varepsilon$ = 3.0, and (dotted 
curve) (20) at $\varepsilon$ = 0.8.
\end{footnotesize}


\begin{thebibliography}{99}
\bibitem{1}
SELEX Collab., M. Mattson {\it et al.}, Phys. Rev. Lett.
{\bf 89}, 112001 (2002).
\bibitem{2}
V.V. Kiselev and A.K. Likhoded, Usp. Fiz. Nauk  {\bf 172}, 497 (2002).
\bibitem{3}
A.V. Berezhnoy, V.V. Kiselev, and A. K. Likhoded, Phys. At. Nucl. {\bf 59}, 870
(1996); S.P. Baranov, Phys. Rev. D {\bf 54}, 3228 (1996);
S.P. Baranov, Phys. Rev. D {\bf 56}, 3046 (1997); 
A.V. Berezhnoy, V.V. Kiselev, A.K. Likhoded, and
A. I. Onichshenko, Phys. At. Nucl. {\bf 60}, 1875 (1997);
A.V. Bereznoy, V.V. Kiselev, A.K. Likhoded, and A.I. Onishchenko,
Phys. Rev. D {\bf 57}, 4385 (1997).
\bibitem{4}
S.P. Baranov and V.L. Slad, Phys. At. Nucl. {\bf 66}, 1730 (2003); 
hep-ph/0602122.
\bibitem{5}
C.-H. Chang, Nucl. Phys. B {\bf 172}, 425 (1980);
R. Baier and R. R\"uckl, Phys. Lett. B {\bf 102}, 364 (1981);
D. Jones, Phys. Rev. D {\bf 23}, 1521 (1981).
\bibitem{6}
R.E. Prange, Phys. Rev. {\bf 110}, 240 (1958).
\bibitem{7}
E. Bagan, H.G. Dosch, P. Godzinsky, S. Narison, and J.-M. Richard,
Z. Phys. C {\bf 64}, 57 (1994).
\bibitem{8}
A. Pukhov {\it et al}., hep-ph/9908288.
\bibitem{9}
C. Peterson, D. Schlatter, I. Schmitt, and P. M. Zerwas, Phys. Rev. D
{\bf 27}, 105 (1983).
\bibitem{10}
P. Collins and T. Spiller, J. Phys. G {\bf 11}, 1289 (1985).
\bibitem{11}
V.D. Kartvelischvili, A.K. Likhoded, and V.A. Petrov,
Phys. Lett. B {\bf 78B}, 615 (1978).
\bibitem{12}
B. Andersson, G. Gustafson, and B. S\"{o}derberg,
Z. Phys. C {\bf 20}, 317 (1983).
\bibitem{13}
ARGUS Collab., H. Albrecht {\it et al}., Phys. Lett. B {\bf 207}, 109 (1988); 
{\bf 247}, 121 (1990).
\bibitem{14}
Particle Data Groups, D. E. Groom {\it et al.}, 
Eur. Phys. J. C {\bf 15}, 1 (2000).
\bibitem{15}
OPAL Collab., G. Alexander {\it et al}., Phys. Lett. B {\bf 364}, 93 (1995);
OPAL Collab., G. Abbiendi {\it et al}., hep-ex/0210031.
\end{thebibliography}
\end{document}